\documentclass[12pt]{article}

\begin{document}
\baselineskip = 0.6 truecm plus 2pt

\begin{titlepage}
\thispagestyle{empty}
\setlength{\topmargin}{1.0cm}
\setlength{\footskip}{1.0cm}
\begin{flushright}
  {\large  hep-th/09114557}
\end{flushright}
\vspace{1.0cm}

\begin{center}
{\Large{\bf D-brane orbiting NS5-branes}}\\

\vspace{1.0cm}
{\large Gyeong Yun Jun\footnote{E-mail: gyjun@ks.ac.kr} and Pyung Seong Kwon\footnote{E-mail: bskwon@ks.ac.kr}}\\

\vspace{0.3cm}
{\large {\it Department of Physics, Kyungsung University,}}\\
{\large {\it 110-1 Daeyeon-dong, Nam-gu, Pusan 608-736, Korea}}
\end{center}

\vspace{1.5cm}
\begin{center}
{\bf Abstract}
\end{center}
{\normalsize  We study real time dynamics of a Dp-brane orbiting a stack of NS5-branes. It is generally  known that a BPS D-brane moving in the vicinity of NS5-branes becomes unstable due to the presence of tachyonic degree of freedom induced on the D-brane. Indeed, the D-brane necessarily falls into the fivebranes due to gravitational attraction and eventually collapses into a pressureless fluid. Such a decay of the D-brane is known to be closely related to the rolling tachyon problem. In this paper we show that in special cases the decay of D-brane caused by gravitational attraction can be avoided. Namely for certain values of energy and angular momentum the D-brane orbits around the fivebranes, maintaining certain distance from the fivebranes all the time, and the process of tachyon condensation is suppressed. We show that the tachyonic degree of freedom induced on such a D-brane really disappears and the brane returns to a stable D-brane.}

\vfill
\begin{center}
\end{center}
\end{titlepage}

\newpage

After the D-brane was found in 1995 \cite{01}, it has been generally conjectured  that our universe may be a stack of D-branes with standard model (SM) fields living on it \cite{02}. But recently there was an argument that true background p-brane immanent in our spacetime may be an NS-NS type brane, rather than D-brane \cite{03}.
Indeed, brane world models including NS-brane have been already considered in the literature \cite{04,05} including "Little String Theory"(LST) \cite{06}. In these models the NS-branes usually appear as background branes near which the D-brane is to be placed, and  in particular in \cite{05} it was argued that these NS-branes play an important role in the context of the cosmological constant problem. In \cite{05} it was shown that in the presence of the background NS-branes the disturbance of the bulk geometry due to quantum fluctuations of SM-fields with support on the D-brane (SM-brane) is highly suppressed in the limit $g_s \rightarrow 0$. So the bulk geometry, as well as the flat intrinsic geometry  of the brane, is practically insensitive to the quantum fluctuations in this limit.

Apart from this, Kutasov noticed \cite{07} that the real time dynamics of D-brane near NS5-branes is closely related to the rolling tachyon problem of the unstable D-brane.
In  \cite{07} he considered a BPS Dp-brane propagating at some distance from a stack of {\it k} parallel NS5-branes of the type II string theory.
In this configuration the supersymmetry of the system is completely broken and the D-brane becomes unstable. Indeed, since the D-brane  experiences an attractive force it either escapes to infinity after deflected by the fivebranes, or it moves towards the fivebranes and eventually decays into a pressureless fluid. Such a decay of the Dp-brane is described by the rolling tachyon solution where the role of the tachyon is played by the radial mode on the D-brane. So, as the D-brane approaches the fivebranes tachyon condensation occurs, and the D-brane turns into some "tachyon matter" state which has an equation of state of a pressureless fluid. Similar configurations have been considered since then by some others \cite{08}, they all obtained basically the same result. They did not find solutions corresponding to a D-brane in orbit around the fivebranes.

If there does not really exist the solution in which the D-brane neither escapes to infinity nor falls into the fivebranes, it would be unnatural to consider the brane world models where D-branes are placed near background NS5-branes, because in the former case the bound state  of the D-brane and the fivebranes can not form,  while in the later case the D-brane will be absorbed into the fivebranes and eventually lose most of its properties including energy, charge and supersymmetry. So it would be interesting if we can find a solution where the D-brane is in orbit around the five-branes. But with the given configuration the solution with stable orbits does not exist (unless we compactify one of the transverse directions), because in the 4d transverse space the D-brane experiences a gravitational potential $ V \sim -{1}/{r^2}$ and this potential does not allow for stable orbits. However, though the solution with stable orbits does not exist, the solution with metastable orbits surely exists. For certain values of energy and angular momentum of the D-brane the attractive force between D-brane and five branes vanishes for all $r$ just as in the case of two parallel BPS D-branes.
 In this paper we will first show that the Dirac-Born-Infeld (DBI) action describing a Dp-brane moving in the vicinity of NS5-branes really admits such a solution with metastable orbits, then we will show that in this case the tachyon potential becomes flat and the tachyon induced on the D-brane turns into a trivial massless constant field. Namely the tachyonic degree of freedom disappears and the D-brane returns to the stable brane.

Before we start we will briefly review the calculations in \cite{07} which are necessary to develop our discussion.
In the presence of {\it k} coincident NS5-branes, the metric, dilaton and NS-NS 3-form fields are respectively given by
\begin{eqnarray}
& &ds^2 =dx_{\mu}dx^{\mu} + H(x^n)dx^{m} dx^{m}\equiv G_{MN} dx^{M} dx^{N}, \nonumber\\
& &e^{2(\Phi-\Phi_0 )}=H(x^n),\\
& &H_{mnp}=-\epsilon^q _{mnp} \partial _q \Phi , \nonumber
\end{eqnarray}
where $x^{\mu} (\mu = 0,1,...5)$ are the coordinates along the world volume of the {\it k} coincident NS5-branes, while $x^{m}\; (m=6,7,8,9)$ the coordinates along the transverse dimensions. Also $H(x^n)$ is a harmonic function
\begin{equation}
 H=1 +{kl^2_s}/{r^2},
\end{equation}
where $r^2 =\sum_{n=6}^{9} x^m x_m$, and $l_s$ is the string length.

Now consider a Dp-brane stretched along the directions $(x_1 , ... x_p)$ with $p\leq 5$, and label the world volume of the D-brane by $\xi^{\mu}$ , $\mu = 0,1,...p$. Then in the static gauge we have $\xi^{\mu}=x^{\mu}$. The dynamics of the world volume fields of the Dp-brane propagating in the above background fields is governed by DBI (Dirac-Born-Infeld) action
\begin{equation}
S_p = - \tau_p \int d^{p+1}\xi e^{-(\Phi -\Phi_0)} \sqrt{-det\mid G_{\mu\nu} +B_{\mu\nu}\mid }\; ,
\end{equation}where $G_{\mu \nu}$ and $B_{\mu \nu}$ are the pullbacks of $G_{MN}$ and $B_{MN}$:
\begin{equation}
G_{\mu \nu} = \frac{\partial x^{M}}{\partial \xi^{\mu}} \frac{\partial x^{N}}{\partial \xi^{\nu}}G_{MN} ,\;\;\;\;B_{\mu \nu} = \frac{\partial{x^M}}{\partial \xi^{\mu}} \frac{\partial{x^N}}{\partial \xi^{\nu}}B_{MN}\;.
\end{equation}
In (4) $x^{M} =(\xi^{\mu} , x^m) $, and $x^m$ now represent the position of the Dp-brane in the transverse space and give rise to world volume scalars $X^{m}(\xi ^{\mu})$. In this paper we will only consider the spatially homogeneous solutions for which $X^m = X^m (t)$. $G_{\mu \nu}$ then reduces to
\begin{equation}
G_{\mu \nu} = \eta_{\mu \nu} + \delta^0 _{\mu} \delta^0 _{\nu}\; H(X^n)\dot{X}^{m}\dot{X}^m \;.
\end{equation}
 We can also allow for nonzero values of $B_{\mu \nu}$ on the D-brane. But it generally breaks the isotropy of the Dp-brane world volume, and generates off-diagonal components of the metric and the stress tensor. In this paper we will assume that the world volume components of the B-field vanish as in \cite{07}, which implies that the induced B-field in (3) vanishes, i.e., $B_{\mu\nu}=0$ in the given configuration.

With these values of world volume fields, the DBI action (3) becomes
\begin{equation}
S_p =- \tau_p V_p \int dt \sqrt{H^{-1}(X^n)-\dot{X}^m \dot{X}^m}\;,
\end{equation}
where  $V_p$ is the volume of the Dp-brane. The action (6) admits a conserved quantity $T_{\mu\nu}$, the stress-energy tensor of the scalar fields $X^m (t)$.
The nonzero components of  $T_{\mu \nu}$ are given by
\begin{eqnarray}
T_{00}= \tau_{p} \frac{1}{H\sqrt{H^{-1}(X^n) - \dot{X}^m \dot{X}^m}}
\;,\;\;\;
T_{ij}=-\tau_p \; \delta_{ij} \; \sqrt{H^{-1}(X^n)-\dot{X}^m\dot{X}^m}\; ,
\end{eqnarray}
where we have set $V_p$ equal to one. $T_{00}$ in (7) is a conserved energy defined by
\begin{equation}
E=P_n\dot{X}^n-\mathcal{L} \;\; ,
\end{equation}
where the momentum $P_n$ is
\begin{equation}
P_n = \frac{\delta\mathcal{L}}{\delta \dot{X}^{n}}= \tau_p \frac{\dot{X}^n}{\sqrt{H^{-1}(X^n)-\dot{X}^m\dot{X}^m}}\;\; .
\end{equation}
One can check that substituting (9) into (8) gives $T_{00}$ in (7). There is another conserved quantity. If we assume that the Dp-brane moves in the $(x^6 , x^7)$ plane it can be shown that the angular momentum defined by $L= X^6P^7 -X^7 P^6$ is also conserved. For instance, see (24).

Let us introduce polar coordinates defined by $X^6=R \cos \theta$ and $X^7=R \sin \theta$.
In these coordinates the angular momentum and the conserved energy take the forms
\begin{equation}
 L=\tau_p\frac{R^2\dot{\theta}}{\sqrt{H^{-1}(R)-(\dot{R}^2+R^2\dot{\theta}^2)}},
\end{equation}

\begin{equation}
E=\tau_p\frac{1}{H\sqrt{H^{-1}(R)-(\dot{R}^2+R^2\dot{\theta}^2)}}\;.
\end{equation}
Solving these two equations in terms of $\dot{R}^2$ and $\dot{\theta}^2$ one obtains
\begin{equation}
 \dot{R}^2=\frac{1}{\epsilon^2 H^2}\Big[\epsilon^2 H-
 (1+\frac{l^2}{R^2})\Big] \;\; ,
\end{equation}
 and
\begin{equation}
 \dot{\theta}^2 =\frac{1}{R^4H^2}\frac{l^2}{\epsilon^2} \;\; ,
\end{equation}
where $l$ and $ \epsilon $ are defined by
$l \equiv L/\tau_{p}$  and $\epsilon \equiv E /\tau_p $, respectively.
 Also from (7) and (11) (and using $E/\tau_p = \epsilon$) one finds
\begin{equation}
T_{ij}= -\frac{1}{H(R)}\; \frac{\tau_p}{\epsilon}\; \delta_{ij} \; .
\end{equation}
The radial equation of motion (12) describes a particle moving in a one dimensional potential
\begin{equation}
V_{eff} = - \frac{1}{\epsilon^2 H^2}\Big[\epsilon^2 H-(1+\frac{l^2}{R^2})\Big]
\end{equation}
with zero energy.

As discussed in \cite{07}, $V_{eff}$ in (15) does not allow for stable orbits, and in general the D-brane escapes to infinity or falls into the fivebranes at late times.
Indeed the author considered two regimes $E>\tau_p$ and $E<\tau_p$, and he found no solutions corresponding to a D-brane in orbit around the fivebranes.
In the intermediate regime, however, there exist solutions in which the D-brane only orbits around the fivebranes without escaping to infinity or falling into fivebranes. Note that $V_{eff}$ in (15) vanishes for all $R$ if

 \begin{equation}
 \epsilon = 1 \longleftrightarrow  E= \tau_p \;,
 \end{equation}
 and
 \begin{equation}
 l=\sqrt{k}l_s \longleftrightarrow  L=\sqrt{k}l_s \tau_p \;\; .
 \end{equation}
 Since $V_{eff}$ vanishes, there is no force that pushes the D-branes into infinity or pulls it  to the fivebranes. The D-brane can maintain its orbit (or the distance) around (from) the fivebranes all the time. This is very reminiscent of the system consisting of two parallel BPS D-branes, where the force between two BPS D-branes precisely vanishes.
So in our system we can place the D-brane at any distance from the fivebranes as we want as in the case of the two parallel BPS D-branes.

Since the D-brane can maintain its orbit, certain amount of $T_{ij}$ is preserved depending on the distance $R$ from the fivebranes. From (14) and (16), $T_{ij}$ is now given by
\begin{equation}
T_{ij}=-\frac{\tau_p}{H(R)} \;\delta_{ij}\;\; .
\end{equation}
Thus  if $R \geq \sqrt{k}l_s$ for instance, then $\mid T_{ij} \mid \geq \tau_p/2$, meaning that more than a half of $T_{ij}$ is preserved if the radius of the orbit is greater than the string scale. This suggests that in our case the D-brane does not decay into a pressureless fluid. Rather, it will have nonzero (negative) pressure unlike the other conventional cases.
Besides this, one also finds from (13) that (16) and (17) imply
\begin{equation}
\dot{\theta}=\frac{\sqrt{k}l_s}{R^2H}\equiv \omega(R)\;\; .
\end{equation}
In (19) the angular velocity takes the value $\omega (R)\rightarrow 0$ as $ R\rightarrow \infty $, while $\omega (R) \rightarrow \frac{1}{\sqrt{k}l_s}$ as $ R\rightarrow 0$.  This is rather unexpected result because it implies that the tangential velocity $v_t(\equiv R\dot{\theta})$ of the  D-brane becomes $ v_t\rightarrow 0$ as $ R\rightarrow 0$.
 Typically the velocity of the particle goes to infinity as the radius of the orbit goes to zero.

  As mentioned already, real time dynamics of the D-brane near NS5-branes necessarily leads to a decay of the D-brane in the ordinary circumstances. The D-brane rolls down to the fivebrane throat  due to gravitational attraction, and as it approaches the fivebrane it loses most of its energy and finally turns into a pressureless fluid. Such a decay of the D-brane is closely related to tachyon condensation on the D-brane, which is described by the rolling tachyon solutions.
 In our case the tachyonic degree of freedom arises from the radial motion of the D-brane moving in the vicinity of NS5-branes. In \cite{07}, it was noticed that $R$ is identified with geometrical tachyon $T$ by the equation.
\begin{equation}
dT =\sqrt{H(R)} dR \;\;\;,
\end{equation}
and the tachyon potential is given by $V(T) =\tau_p/\sqrt{H(R(T))}$.  From (20) one finds that as $R\rightarrow 0$, $T(R) \sim \sqrt{k}l_sln R/\sqrt{k}l_s$ or $R(T)\sim\sqrt{k}l_s e^{T/\sqrt{k}l_s}$, and therefore $V(T)/\tau_p \sim 1/\sqrt{H(R(T))}\sim e^{T/\sqrt{k}l_s}$, indicating that the potential $V(T)$ goes exponentially to zero as $R\rightarrow 0$. This precisely coincides with the behavior exhibited by the tachyon potential relevant for the rolling tachyon solutions. So the tachyon rolls towards the minimum of the potential, and as a result the D-brane collapses into a pressureless fluid.
In our case, however, this is not to be the case anymore. Since the D-brane maintains a certain distance from the fivebranes, the process of tachyon condensation is expected to be suppressed. Indeed, $\dot{R}=0$ implies $\dot{T}=0$ from (20), which means that tachyon does not roll anymore in our case, i.e., when $\epsilon=1$ and $l=\sqrt{k}l_s$.

The fact that  tachyon does not roll when $\epsilon=1$ and $l=\sqrt{k}l_s$ can be understood as follows. First, rewrite the DBI action (6) in terms of $\dot{R}$ and $\dot{\theta}$ :
\begin{equation}
S_p =- \tau_p \int dt \sqrt{ H^{-1}(R) - (\dot{R}^2+R^2\dot{\theta}^2)} \;\; ,
\end{equation}
where we have set $V_p=1$ as before. The equations of motion for $R$ and $\theta$ are then respectively given by
\begin{equation}
\frac{d}{dt}\Bigg(\frac{\dot{R}}{\Sigma(\dot{\theta})}\Bigg)=\frac{R}{\Sigma(\dot{\theta})}\;[\;\dot{\theta}^2-\omega^2(R)\;] \;,
\end{equation}
where
\begin{equation}
\Sigma(\dot{\theta})\equiv \sqrt{H^{-1}(R)-(\dot{R}^2+R^2\dot{\theta}^2)} \;,
\end{equation}
and
\begin{equation}
\frac{dL}{dt}=0
\end{equation}
with $L$ given by (10). (Check that $\dot{R}=0$ with $\epsilon=1$ and $l=\sqrt{k}l_s$ becomes a solution to (22). Also (24) ensures that $L$ in (10) is a constant of motion.) As can be seen from (20) the geometrical tachyon only arises from the radial mode $R$.  So if we want to analyze the tachyonic behavior it is only enough to consider the radial motion of the D-brane.
(22), however, contains both $\dot{R}$ and $\dot{\theta}$. To express it only in terms of $\dot{R}$ and $R$, we solve (10) for $\dot{\theta}$ to get
\begin{equation}
\dot{\theta}^2=\frac{l^2}{R^4 H^2}\frac{H}{(1+\frac{l^2}{R^2})}-\frac{\dot{R}^2}{R^4}\frac{l^2}{(1+\frac{l^2}{R^2})} \;,
\end{equation}
and also using (25) we obtain
\begin{equation}
\Sigma(\dot{\theta})=\frac{\Sigma(0)}{\sqrt{1+\frac{l^2}{R^2}}} \;,
\end{equation}
where
\begin{equation}
\Sigma(0) = \sqrt{H^{-1}(R)-\dot{R}^2} \;.
\end{equation}
Substituting (25) and (26) into (22) one finally obtains a $\dot{\theta}$-independent equation of motion
\begin{equation}
\frac{d}{dt}\Bigg[ \frac{\sqrt{1+\frac{l^2}{R^2}}}{\Sigma(0)} \;\dot{R}\Bigg]
=\frac{1}{\sqrt{1+\frac{l^2}{R^2}}} \frac{R}{\Sigma(0)}\Bigg[\Bigg(\frac{l^2}{kl_{s}^{2}}-1\Bigg)\omega^2(R) -l^2 \frac{\dot{R}^2}{R^4}\Bigg] \;\; .
\end{equation}

Now consider an action of the form
\begin{equation}
\tilde{S_p}=-\tau_p \int dt \;  \sqrt{1+\frac{l^2}{R^2}}\;\sqrt{H^{-1}(R)-\dot{R}^2} \;\; .
\end{equation}
 Since (29) does not contain $\dot{\theta}$ term it only gives an equation of motion for $R$, and one can show that the equation of motion following from (29) precisely coincides with (28). Besides this, the conserved energy obtained  from (29) is given by
\begin{equation}
\tilde{E}=\tau_p\frac{\sqrt{1+\frac{l^2}{R^2}}}{H\sqrt{H^{-1}(R)-\dot{R}^2}},
\end{equation}
which is also equal to $E$ in (11) due to (26):
\begin{equation}
\tilde{E}=E=\tau_p\epsilon\;\;.
\end{equation}
These things indicate that $\tilde{S_p}$ is classically equivalent to the original action $S_p$ as far as the radial motion is concerned, which in turn means that the tachyonic behavior described by $\tilde{S_p}$ is equivalent to that described by the original action $S_p$ because the tachyon $T$ is only a field redefinition of $R$ (Eq.(20)).
Indeed the one dimensional motion described by $\tilde{S}_p$ exactly coincides with the radial motion described by $S_p$.

Let us now rewrite $\tilde{S}_p$  in terms of $T$:
\begin{equation}
\tilde{S}_p =- \int dt \tilde{V}(T)\sqrt{1-\dot{T}^2} =\int dt L(t)\;\;\;,
\end{equation}
where  $\tilde{V}(T)$  is given by
\begin{equation}
\tilde{V}(T) = \tau_p  \frac{\sqrt{1+\frac{l^2}{R^2}}}{\sqrt{H(R(T))}}  \;\;.
\end{equation}
We observe that the tachyon potential has been changed from $V(T)=\tau_p/\sqrt{H(R(T))}$ into (33). Such a change of the tachyon potential is obviously due to the orbiting motion of the D-brane.
Note that the change has occurred in compensation for the elimination of the $\dot{\theta}$ term.
The tachyon potential $\tilde{V}(T)$ has a remarkable feature. For $l=\sqrt{k}l_s$, it simply becomes a constant:
\begin{equation}
\tilde{V}(T)=  \tau_p \;\;.
\end{equation}
The decay of unstable D-brane essentially occurs as the tachyon rolls towards the minimum of the tachyon potential. But in the case  $l=\sqrt{k}l_s$, the tachyon does not roll (provided the initial condition $\dot{T}=0$ is met) because the potential $\tilde{V}(T)$ is flat, and the decay of the D-brane is necessarily suppressed. To be more precise, in the case $l=\sqrt{k}l_s$ the geometrical tachyon induced on the D-brane turns into a trivial massless constant field, meaning that the tachyonic degree of freedom on the D-brane disappears and the D-brane returns to the stable brane.  Thus the whole issue regarding  rolling tachyon, including gravitational radiation, becomes irrelevant to this case regardless of whether it is considered at the tree level or quantum level.

The fact that the unstable D-brane returns to the stable brane when $l=\sqrt{k}l_s$ and $\epsilon=1$ can be confirmed as follows.
In the case of the usual unstable D-brane (of the bosonic string theory) the spatially homogeneous tachyon is often described by
\begin{equation}
T(X^0)=\lambda \cosh X^0
\end{equation}
and the corresponding tachyon potential
\begin{equation}
V(T) =\frac{\tau_p}{\coth\frac{T}{2}}\;\;\;.
\end{equation}
The tree level analysis of the boundary conformal field theory (BCFT) shows \cite{09} that the energy-stress tensor $T_{\mu\nu}$ corresponding to the above tachyon profile takes the form
\begin{equation}
T_{00}=\frac{\tau_p}{2}(1+\cos2\pi \lambda)\; ,\;\;\;\;\;\; T_{ij}=-\tau_p\;f(t)\;\delta_{ij}\;,
\end{equation}
where $t=x^0$ and $f(t)$ is given by
\begin{equation}
f(t)=\frac{1}{1+\sin(\lambda \pi)e^{t}} + \frac{1}{1+\sin(\lambda\pi)e^{-t}}-1 \;\;.
\end{equation}
For positive $\lambda$ the function $f(t)$ goes to zero as $t\rightarrow \infty$, showing that the pressure vanishes asymptotically and the D-brane decays into a pressureless fluid. This is what happens as the tachyon rolls towards the minimum of the potential $V(T)$.

 In our case, however, the tachyon does not roll as mentioned already.
Consider the equation of motion following  from (32):
\begin{equation}
\dot{T}=\sqrt{1-\frac{\tilde{V}^2}{\tilde{E}^2}}\;\;.
\end{equation}
 Using (31), (34) and (39) one finds that $\dot{T}$ really vanishes when $l=\sqrt{k}l_s$ and $\epsilon=1$.   Since the tachyon does not roll,  the function $f(t)$ is expected to be a constant. There is a simple way to find $f(t)$ which does not use the BCFT analysis. On general grounds the function $\tau_p f(t)$, which is a partition function on the disk in BCFT, can be identified as the on-shell value of $-L(t)$ \cite{10}.
Substituting (34) together with $\dot{T}=0$ into $L(t)$ one finds that the function $f(t)$ is just equal to one:
\begin{equation}
f(t)=1\;\;\;,
\end{equation}
and consequently the components of $\tilde{T}_{\mu\nu}$ following from $\tilde{S}_p$ become
\begin{equation}
\tilde{T}_{00}=\tilde{E}= \tau_p \;,\;\;\;\;\;\;  \tilde{T}_{ij}=-\tau_p\delta_{ij}\;\;\;,
\end{equation}
which are  typical of the BPS D-brane. So we expect that the  action $S_p$ (with $\epsilon=1$ and $l=\sqrt{k}l_s$ ) also describes a BPS D-brane because the tachyonic behavior described by $\tilde{S}_p$ is equivalent to that described by $S_p$.

 The D-brane described by $S_p$ really appears as a BPS brane to an observer living on that D-brane. Notice that the conditions $\epsilon=1$ and $l=\sqrt{k}l_s$ imply $R=$ const $\equiv R_0$ and $\dot{\theta}=\sqrt{k}l_s/HR^2$, which then gives $G_{00}=-1/H(R_0)$ and $G_{ij}=\eta_{ij}$ from (5). Thus the observer on the D-brane finds the stress tensor
 \begin{equation}
 T_{\mu\nu}^{(brane)}=\; -\hat{\tau_p}\eta_{\mu\nu}\;,\;\;\;\;\; \hat{\tau_p}\equiv \frac{\tau_p}{H(R_0)}\;,
 \end{equation}
 and sees the geometry $ds^2=-d\tau^2 + d\vec{x}_p^2$, where the proper time $\tau$ is related to $t$ by $d\tau^2 =- G_{00}dt^2$. (42) is precisely the stress tensor of the BPS D-brane with a tension $\hat{\tau}_p$.  Thus the D-brane  orbiting  around NS 5-branes with $\epsilon=1$ and $l=\sqrt{k}l_s$ appears  as a BPS D-brane with  an effective tension $\hat{\tau}_p$ (and with no NS 5-branes nearby) to an observer on the D-brane. This suggests that the effects of the NS 5-branes on the D-brane have disappeared  due to the orbiting motion of the D-brane. The presence of the NS5-branes only changes the tension  measured by an observer on the D-brane. The effective tension $\hat{\tau}_p$ goes to zero as $R_0\rightarrow 0 $, while it approaches $\tau_p$ as $R_0 \rightarrow\infty$. In the brane world cosmology $\hat{\tau}_p$ serves for a cosmological constant and makes a contribution to the dark energy.

The whole analysis of this paper has been made at the classical level.
But even at the quantum level we do not need to be concerned about the gravitational or closed string radiations generated by tachyon because the tachyon field coupled with graviton or closed string modes has been disappeared already. The aim of this paper is to examine the possibility of avoiding the Kutasov's conjecture, which also has been made at the classical level, that a D-brane moving around NS5-branes is necessarily absorbed into the fivebranes and eventually decays into a "pressureless fluid". According to the result of this paper the decay of the D-brane can be avoided if the brane has specific values of energy and angular momentum.  In that case the D-brane orbits around the fivebranes, maintaining  certain distance from the fivebranes all the time, and consequently a stable bound state of D-brane and fivebranes can be formed. But before concluding, it should be mentioned that the discussion of this paper was based on the assumption that the radiation emitted from the D-brane is entirely generated by the tachyonic degree of freedom induced on the brane, as it was
in other papers including \cite{07}. In general the D-brane is regarded as a source for the closed string modes. It couples to the metric, dilaton and the $(p+1)$-form R-R gauge field of the type II string theories. So if the D-brane is accelerated, or rotating for instance, one would expect it to emit Larmor-type radiation of these fields. Though this is not the main point of this paper it may be necessary to address it for the completeness of our discussion.

In \cite{11}, the radiation emitted by accelerating D-branes has been studied in the linear approximation to the supergravity limit of the string theory. Assuming that the D-brane is moving in three uncompactified spatial dimensions the authors found that the total radiated power per unit mass (energy) of the D-brane is given by $P/M\sim \kappa^2 M |\dot{\vec{v}}|^2$, where $\kappa^2 \equiv\kappa^2_{10} /V_6$ with $V_6$ the volume of the extra dimensions is the 4d gravitational coupling, while $M$ and $\dot{\vec{v}}$ are the mass and the acceleration of the  Dp-brane respectively. This result, however, may not be directly applicable to our case because in our case the spatial dimensions of the spacetime in which the radiation propagates are four, instead of three, and the green function is therefore proportional to $\sim 1/r^2$, instead of $\sim 1/r$, where $r$ represents the distance from the source (the rotating Dp-brane) to the observation point of the radiation and it characterizes the scale of the transverse dimensions. Thus the Poynting vector is expected to fall off as $\sim 1/r^4$, and we estimate $P/M$ to be $P/M \sim\kappa^2 M | \dot{\vec{v}}|^2 /r$ where $\kappa^2$ is now given by $\kappa^2 =\kappa^2_{10} / V_5$. We see that $P/M$ goes to zero as $r\rightarrow \infty$.

Apart from this, there is a good reason for neglecting the energy loss resulting from the Larmor-type radiation. In order to see this, consider the case $p=5$ for instance. For $p=5$, $\kappa^2$ and $M$ are respectively given by $\kappa^2\sim g_s^2 \alpha'^4 /V_5$, $M\sim V_5/g_s \alpha'^3$, and since $|\dot{\vec{v}}| \sim R_0\dot{\theta}^2$, $P/M$ becomes\footnote[1]{If the D5-brane is wrapped on a 2-cycle to become an effective D3-brane, $P/M$ gets even smaller than (43). In this case $P/M$ acquires an extra factor $V^{(c)}_2/V_2$, where $V_2^{(c)}$ is the volume of the 2-cycle, while $V_2$ the volume of the 2d uncompactified space.}
\begin{equation}
\frac{P}{M} \sim \bigm(\frac{g_s}{k}\Bigm)\Bigm(\frac{1}{r}\Bigm)\frac{(\sqrt{k}l_s/R_0)^6}{[1+(\sqrt{k}l_s/R_0)^2]^4}\;\;.
\end{equation}
(43) is only valid for $R_0 \gg \sqrt{k}l_s$ because it has been obtained by assuming the situation in which the accelerated brane moves in a background which is almost a flat spacetime. Also in our discussion it has been assumed that the 4d transverse space is noncompact\footnote[2]{In general the green function describing radiation emitted from a single source does not exists in a compact space, and therefore the Larmor-type radiation is automatically suppressed in this case.} as in \cite{07}, and we may take $r\sim\infty$ which characterizes the scale of the transverse dimensions.

For these reasons we are only allowed to consider the case $r\gg R_0\gg\sqrt{k}l_s$ as far as the Larmor-type radiation is concerned. Clearly, the ratio $P/M $ in (43) goes to zero in the limit $g_s \rightarrow 0$, which suggests that the Larmor-type radiation could be ignored in our case. Indeed, the time it takes for the whole energy of the D-brane to be dissipated away by the Larmor-type radiation is given by $t=(1/c)M/P$, which is expected to be very large for $g_s \rightarrow 0$ and $r\gg R_0 \gg\sqrt{k}l_s$ . In fact, as long as $r$ or the size of the transverse dimensions is not so small, $t$ can be arbitrarily large in the limit $g_s \rightarrow 0$ depending on how large the ratio $R_0/\sqrt{k}l_s$ is. In any case, however, if $r$ and $R_0$ are not sufficiently large, we may have to resort to the quantum theory to avoid the Larmor-type radiation as in the case of an electron orbiting the proton where the electron is not absorbed into the proton due to orbit quantization.

\vskip 1cm
\begin{center}
{\Large \bf Acknowledgements}\\
\end{center}

This work is supported by Kyungsung University in 2009.

\end{document}